\documentclass[12pt]{article}

\usepackage[utf8]{inputenc}
\usepackage{latexsym, graphicx} 
\usepackage{amsmath,amsfonts,amssymb}
\usepackage{mathrsfs}
\usepackage{tensor}
\usepackage{cite}
\usepackage[colorlinks=true,linkcolor=blue,citecolor=blue,urlcolor=blue,filecolor=black]{hyperref}

\renewenvironment{thebibliography}[1]{
  \begin{oldthebibliography}{#1}
    \setlength{\itemsep}{2pt}
    \setlength{\parskip}{2pt}
}
{
  \end{oldthebibliography}
}

\numberwithin{equation}{section} 

\newcommand*{\scri}{\ensuremath{\mathscr{I}}}

\newcommand*{\dd}{\mathop{}\!d}

\newcommand{\zb}{\bar{z}}
\newcommand{\wb}{\bar{w}}
\def\p{\partial}
\newcommand{\soft}{\big|_{\text{soft}}}
\newcommand{\Soft}{\Big|_{\text{soft}}}
\newcommand{\cc}{\text{c.c.}}

\begin{document}

\begin{titlepage}
  \thispagestyle{empty}

  \begin{flushright}
  \end{flushright}

\vskip3cm

\begin{center}  
{\Large\textbf{Celestial soft dressings from generalised Wilson lines}}

\vskip1cm

\centerline{Kevin Nguyen$^\dagger$\footnote{kevin.nguyen@kcl.ac.uk}, Alan Rios Fukelman$^\dagger$\footnote{alan.rios\_fukelman@kcl.ac.uk} and Chris D. White$^\ddag$\footnote{christopher.white@qmul.ac.uk}}

\vskip1cm

{\it{$^\dagger$Department of Mathematics, King's College London,\\
The Strand, London WC2R 2LS, United Kingdom}}\\
\vskip 0.5cm
{\it{$^\ddag$Department of Physics and Astronomy, Queen Mary University of London,\\
327 Mile End Road, London E1 4NS, United Kingdom}}\\
\vskip 0.5cm

\end{center}

\vskip1cm

\begin{abstract} 
In this review article, we revisit the connection between dressing of scattering states in quantum electrodynamics by clouds of soft photons, and their dressing by (generalised) Wilson line operators. In particular, we show that the leading and subleading soft conformal dressings considered in the context of celestial holography can be straightforwardly obtained from  generalised Wilson lines, and that this only requires knowledge of the asymptotic behaviour of the photon field near null and timelike infinity.  
\end{abstract}

\end{titlepage}

{\hypersetup{linkcolor=black}
  \tableofcontents
  }
  
\section{Introduction}
\label{sec:intro}
In quantum electrodynamics (QED) and other gauge theories, the traditional scattering S-matrix between Fock states infamously suffers from infrared divergences \cite{Mott,Bloch:1937pw,Yennie:1961ad,Kinoshita:1962ur,Lee:1964is,Weinberg:1965nx}. In order to overcome this issue, dressing of Fock states by coherent states of soft photons, commonly referred to as Faddeev--Kulish (FK) dressing, was designed such that S-matrix elements be free of these infrared divergences \cite{Chung:1965zza,Kibble:1968sfb,Kibble:1968oug,Kibble:1968npb,Kibble:1968lka,Kulish:1970ut}. A detailed account of this and other related topics is provided in \cite{Agarwal:2021ais}. Importantly, a close relation between these dressed states and the gauge-invariant quantum fields constructed through the introduction of Wilson lines \cite{Dirac:1955uv,Mandelstam:1962mi} was also 
later uncovered \cite{Zwanziger:1973mp,Jakob:1990zi}. 

More recently it has been observed that many infrared features of the S-matrix are actually controlled by large gauge transformations (LGT), i.e., gauge transformations with support at infinity \cite{Strominger:2017zoo,Strominger:2013lka,He:2014cra,Kapec:2015ena,Campiglia:2015qka,Lysov:2014csa,Gabai:2016kuf,Kapec:2017tkm}. In particular it has been understood that FK states are eigenstates of the LGT charge, and that the latter characterises scattering superselection sectors \cite{Gabai:2016kuf,Kapec:2017tkm,Choi:2017ylo}. In fact conservation of the LGT charge is equivalent to the leading soft photon theorem \cite{Strominger:2013lka,He:2014cra,Kapec:2015ena,Campiglia:2015qka}, which trivialises when considering the scattering of dressed states rather than that of Fock states. Given that there also exists a subleading soft photon theorem \cite{Low:1954kd,Gell-Mann:1954wra,Low:1958sn,Burnett:1967km}, a natural extension was the construction of dressed states that trivialise this subleading soft theorem as well \cite{Choi:2019rlz}. However the subleading dressings do not relate to infrared divergences and are therefore not strictly needed. In this article we observe that the subleading soft dressing can be given a different interpretation in terms of a generalised Wilson line operator, introduced in the literature to efficiently compute scattering amplitudes at subleading order in a soft expansion~\cite{Laenen:2008gt,Laenen:2010uz,White:2014qia,Bahjat-Abbas:2019fqa,White:2011yy,Bonocore:2015esa,Bonocore:2016awd}. 

Largely motivated by the connection between soft theorems and LGT charge conservation, a new approach to scattering amplitudes emerged under the name of \textit{celestial holography}. For recent reviews of this rapidly growing field we refer the reader to \cite{Pasterski:2021raf,McLoughlin:2022ljp}. In that approach a different basis of one-particle states is used, given by the set of conformal primary states that are boost rather than momentum eigenstates \cite{deBoer:2003vf,Cheung:2016iub,Pasterski:2016qvg,Pasterski:2017kqt,Pasterski:2017ylz,Fotopoulos:2019vac,Law:2020tsg,Iacobacci:2020por,Narayanan:2020amh,Pasterski:2020pdk}. These behave precisely as (quasi)-conformal primaries for the Lorentz group $SO(1,3) \simeq SL(2,\mathbb{C})$. In this conformal basis the soft theorems take the form of Ward identities associated with conserved currents of a two-dimensional conformal field theory \cite{He:2015zea,Himwich:2019dug}. On the other hand infrared divergences are then accounted for by a particular conformal field, namely the Goldstone mode of spontaneously broken LGT \cite{Nande:2017dba}. In order to define an S-matrix free of these infrared divergences, a notion of dressing appropriate to the conformal primary states has been similarly proposed \cite{Kapec:2017tkm,Arkani-Hamed:2020gyp,Pasterski:2021dqe}, which is different but closely related to the FK dressing. 

In this article we revisit the connection between (generalised) Wilson line operators and the various notions of dressings discussed above. Our aims are not only to collect useful results in one place, but also to explain the connection between them. In particular we show how the leading and subleading soft dressings are easily reproduced from the Wilson line operators through use of the asymptotic expansions of the gauge potential near null infinity~$\scri$ and timelike infinity~$i^+$. This approach therefore ties together the asymptotic structure of QED with the soft dressings more directly, and gives a nice geometrical picture for the latter in terms of Wilson lines tracking the classical trajectories of the scattered particles.   

The article is organised as follows. We start in section~\ref{sec:preliminaries} by reviewing the leading and subleading FK soft dressings, the conformally soft dressings and the Wilson line dressings, together with the known relations between them. In section~\ref{sec:asymptotics} we describe the asymptotic regions near $\scri$ and $i^+$, as well as the corresponding parametrisation of classical particle trajectories. In appendix~\ref{app:expansion at scri} and \ref{app:expansion at i+} we study the asymptotic structure of the electromagnetic potential near null infinity~$\scri$ and timelike infinity~$i^+$, respectively. We use it in sections~\ref{sec:massless soft dressings} and \ref{sec:massive soft dressings} to derive the soft dressings from the (generalised) Wilson line operators associated with massless and massive scattering states, respectively.

\section{Preliminaries}
\label{sec:preliminaries}
We start by recalling three different versions of dressings and the known relations between them. These are Faddeev--Kulish (FK), conformally soft and Wilson line dressings. We will restrict the discussion to QED for simplicity.

\paragraph{Faddeev--Kulish dressings.} 
As reviewed in the introduction, the FK dressing of charged particle states by clouds of soft photons were introduced in order to define an S-matrix free of infrared divergences. For a one-particle state $|p,J \rangle$ with momentum $p$, helicity $J=\pm s$ and electric charge $e Q$, the corresponding FK dressed state can be written
\begin{equation}
||p,J\rangle\rangle=\tilde W_0\, \tilde W_1\, |p,J \rangle\,,
\end{equation}
with the \textit{leading} FK dressing given by \cite{Kulish:1970ut}
\begin{equation}
\label{leading FK}
\tilde W_0=\exp \left[e Q \int \frac{d^3 \vec q}{(2\pi)^3} \frac{f(\omega)}{2 \omega} \frac{p^\mu}{p \cdot q} \left( a^\dagger_\mu( q)- a_\mu( q) \right)\right]\,,
\end{equation}
and the \textit{subleading} FK dressing given by \cite{Choi:2019rlz}
\begin{equation}
\label{subleading FK}
\tilde W_1=\exp \left[-i e Q \int \frac{d^3 \vec q}{(2\pi)^3} \frac{f(\omega)}{2 \omega} \frac{q_\nu J^{\nu\mu}}{p \cdot q} \left( a^\dagger_\mu( q)+ a_\mu( q) \right)\right]\,.
\end{equation}
In the above expressions $a^\dagger_\mu(q)$ is the creation operator for a photon of null momentum $q^\mu$ with frequency $\omega=|\vec{q}\,|$,
$J_{\mu\nu}$ is the Lorentz generator (or total angular momentum), and $f(\omega)$ a distribution with vanishingly small support around $\omega=0$ satisfying $f(0)=1$. The latter conditions simply mean that the dressing only retains the leading nontrivial contribution in the limit $\omega \to 0$. 

\paragraph{Conformally soft dressings.} This second type of dressing naturally arises in the description of scattering amplitudes in a basis of conformal primary states \cite{Kapec:2017tkm,Arkani-Hamed:2020gyp,Pasterski:2021dqe}. Conformal primary states are typically denoted $|\Delta,J, w,\wb \rangle$, and depend on the conformal dimension $\Delta$ and helicity $J=\pm s$ as well as some insertion point $(w,\wb)$ on the celestial sphere at null infinity. Conformally soft dressings are then the analogue of the FK dressings in this context and are still formally given by the expressions \eqref{leading FK}-\eqref{subleading FK}, however with the important difference that $f(\omega)=1$ across the whole spectrum. This means in particular that both hard and soft photons contribute to the dressing, and that conformally dressed states have infinite energy. Denoting by $W_0$ and $W_1$ the leading and subleading conformally soft dressing operators, we have the relation
\begin{equation}
\label{soft W = tilde W}
W_0\big|_{\text{soft}}=\tilde W_0\,, \qquad W_1\big|_{\text{soft}}=\tilde W_1\,,
\end{equation}
where, given a generic quantity $Q=\int_0^\infty d\omega\, Q(\omega)$ involving photons of all frequencies $\omega$, we define its \textit{soft component} through insertion of a function with vanishingly small support around $\omega=0$ and satisfying $f(0)=1$,
\begin{equation}
\label{definition soft}
Q\big|_{\text{soft}}\equiv \int_0^\infty d\omega f(\omega) Q(\omega)\,.
\end{equation}

The explicit expressions given in \cite{Kapec:2017tkm,Arkani-Hamed:2020gyp,Pasterski:2021dqe} for the conformally soft dressings of a massless state $|\Delta,J,w,\wb \rangle$ are 
\begin{equation}
\label{W0}
W_0=e^{-i Q\, \mathcal{S}(w,\wb)}\,,
\end{equation}
and
\begin{equation}
\label{W1}
W_1=\exp\left[\frac{Q}{\omega} \left(2h\, \partial_w \mathcal{Y}^w+\mathcal{Y}^w \partial_w+2\bar h\, \partial_{\wb} \mathcal{Y}^{\wb}+\mathcal{Y}^{\wb} \partial_{\wb} \right)\right]\,,
\end{equation}
where $\Delta=h+\bar h$ and $J=h-\bar h$.
Here $\mathcal{S}(w,\wb)$ and $\mathcal{Y}^w(w,\wb)$ are conformal fields of weights $h=\bar h=0$, given in terms of photon creation and annihilation operators by \cite{Pasterski:2021dqe}
\begin{align}
\label{S and Y}
\begin{split}
\mathcal{S}(w,\wb)&=\frac{i e}{\sqrt{2}(2\pi)^2} \int_0^\infty d\omega \left[ \partial_{\wb}^{-1} (a_-(q) - a^\dagger_+(q))+\partial_{w}^{-1} (a_+(q) - a^\dagger_-(q)) \right]\,,\\
\mathcal{Y}^w(w,\wb)&=\frac{e}{\sqrt{2}(2\pi)^2} \int_0^\infty d\omega\, \omega\, \partial_{w}^{-2} (a_+(q) + a^\dagger_-(q))\,,\\
\mathcal{Y}^{\wb}(w,\wb)&=\frac{e}{\sqrt{2}(2\pi)^2} \int_0^\infty d\omega\, \omega\, \partial_{\wb}^{-2} (a_-(q) + a^\dagger_+(q))\,,
\end{split}
\end{align}
where there is an implicit parametrisation $q(\omega,w,\wb)$ of the photon momentum, given below in \eqref{null pi}. For massive states the leading conformally soft dressing can also be found in \cite{Kapec:2017tkm,Arkani-Hamed:2020gyp}, however the subleading dressing expressed in terms of conformal fields has yet to be worked out.

\paragraph{Wilson line dressings.}
The leading conformally soft factor $W_0$ can be alternatively understood as a dressing by a Wilson line along the particle classical trajectory \cite{Zwanziger:1973mp,Jakob:1990zi,Choi:2018oel}. One justification for the presence of a Wilson line is simply the construction of gauge-invariant field variables at the cost of introducing some controlled degree of nonlocality \cite{Dirac:1955uv,Mandelstam:1962mi}. Indeed, consider a local quantum field $\phi(x)$ with electric charge $e Q$, which acquires a phase under local $U(1)$ transformations,
\begin{equation}
\phi(x) \mapsto e^{-i eQ \Lambda(x)} \phi(x)\,, \qquad A_\mu(x) \mapsto A_\mu(x) + e\, \partial_\mu \Lambda(x)\,.
\end{equation}
The phase factor can be compensated for by considering a line segment $\Gamma[x_0,x]$ with endpoints $x_0$ and $x$, and dressing the local field $\phi(x)$ with the corresponding Wilson line operator,
\begin{equation}
\phi(x|\Gamma) \equiv e^{i  Q \int_\Gamma A}\, \phi(x)\,.
\end{equation}
The transformation of the dressed field is now instead
\begin{equation}
\label{global phase}
\phi(x|\Gamma) \mapsto e^{-ieQ \Lambda(x_0)} \phi(x|\Gamma)\,.
\end{equation}
Fixing the reference point $x_0$ once and for all (such as the origin of Minkowski space), the above is a global phase shift common to all dressed field variables $\phi(x|\Gamma)$ independently of $x$, and local gauge invariance is therefore achieved. 

The connection with the conformally soft dressing $W_0$ of a single-particle state $|p,J\rangle$ arises when choosing the path $\Gamma$ to be the classical trajectory of the corresponding particle. As is well-known a Wilson line carries infinite energy, and thus corresponds to the dressings $W_{0}$ for which $f(\omega)=1$ across the whole spectrum. Specifically, considering $A_\mu$ to be the \textit{radiation} field\footnote{That part of the gauge potential which is sourced by onshell particles is known to account for Coulomb phases \cite{Jakob:1990zi}. This will not be further discussed here.}, we have the equality \cite{Jakob:1990zi}
\begin{equation}
\label{leading Wilson}
W_0=e^{-i Q \int_\Gamma A}\,.
\end{equation}

This equivalence extends to the subleading dressing as well. Still considering the path $\Gamma$ to be the classical trajectory of the particle, the subleading FK dressing \eqref{subleading FK} can be written\footnote{We fixed the normalisation of the exponent in order to agree with the expression \eqref{W1} in section~\ref{sec:massless soft dressings}.} 
\begin{equation}
\label{subleading Wilson}
W_1=\exp \left[-\frac{iQ}{8} \int_\Gamma  F_{\mu\nu} J^{\mu\nu} \right]\,.
\end{equation}
In the leading soft case, we saw that FK dressings are in turn related to Wilson lines. The question then naturally arises as to whether the subleading FK dressing of eq.~(\ref{subleading Wilson}) can itself be associated with a generalisation of the conventional Wilson line. Indeed this is the case, where the relevant generalised Wilson line was first discussed in refs.~\cite{Laenen:2008gt,Laenen:2010uz}, in the context of collider physics. That reference discussed gluon radiation from scalar particles, and argued that one may indeed write a generalised Wilson line describing radiation up to next-to-soft level, where the additional contribution to the Wilson line involves a contraction of the field strength tensor with the generator of spin transformations, consistent with eq.~(\ref{subleading Wilson}). Further work has established that the full angular momentum generator is indeed obtained from conventional QCD approaches~\cite{White:2014qia}, which must anyway be the case on general grounds (e.g. Lorentz and gauge invariance). Related results in gravity were obtained in ref.~\cite{White:2011yy}, and applications to collider physics may be found in refs.~\cite{Bahjat-Abbas:2019fqa,Bonocore:2015esa,Bonocore:2016awd}. Also note that Wilson line dressings in AdS together with the corresponding flat space limit has been discussed in \cite{Duary:2022afn}.  

\section{Asymptotic regions and classical trajectories}
\label{sec:asymptotics}
We aim to describe the FK dressings in terms of asymptotic components of the photon field. To achieve this we will exploit the relation between soft dressings and (generalised) Wilson lines reviewed in the preceding section.

A distinction between massless and massive fields will be made, as the asymptotic regions where the corresponding wavefunctions are supported are of a different nature. Massless fields and the corresponding classical null rays extend to future (past) null infinity $\scri^+\, (\scri^-)$, while massive fields and the corresponding timelike geodesics extend to future (past) timelike infinity $i^+\, (i^-)$. We present the coordinate systems adapted to $\scri^+$ and $i^+$ and describe some relevant aspects of the corresponding classical trajectories.

\paragraph{Null infinity.}
Retarded coordinates $x=(r,u,z,\zb)$ are best adapted to the description of~$\scri^+$. The relation to cartesian coordinates $X^\mu$ can be conveniently written \cite{Himwich:2020rro}
\begin{equation}
\label{coord transf retarded}
X^\mu=u\, n^\mu+r\, \hat q^\mu(z,\zb)\,,
\end{equation}
where $n^\mu$ and $\hat q^\mu(z,\zb)$ are null vectors with cartesian components given by
\begin{align}
\label{q and n}
\begin{split}
\hat q^\mu(z,\zb)&= \frac{1}{\sqrt{2}}\left(1+z \zb\,, z+\zb\,, -i(z-\zb)\,, 1-z \zb\right)\,,\\
n^\mu&=\frac{1}{\sqrt{2}}\left(1,0,0,-1\right)\,,
\end{split}
\end{align}
and satisfying the useful relations
\begin{equation}
n \cdot \hat q=-1\,, \qquad \hat q_i \cdot \hat q_j=-\, |z_{ij}|^2\,.
\end{equation}
This allows to easily obtain the retarded coordinates $r$ and $u$ from the position vector $X^\mu$ through the following Lorentz contractions,
\begin{equation}
\label{r and u}
r=-n \cdot X\,, \qquad u=-\hat q \cdot X\,.
\end{equation}
In retarded coordinates the flat metric takes the form
\begin{equation}
\label{flat Bondi gauge}
ds^2=\eta_{\mu\nu} \dd X^\mu \dd X^\nu=-2 \dd u \dd r+ 2 r^2 \dd z \dd \zb\,.
\end{equation}
The location of future null infinity $\scri^+$ corresponds to the limit $r \to \infty$. It is a three-dimensional null manifold covered by the coordinates $(u,z,\zb)$.

Let us now turn to the parametrisation of generic null rays passing through a reference point $X_0$ in Minkowski space. The parametrisation of a null ray is simply given by
\begin{equation}
\label{null trajectory}
X^\mu(s)=X_0^\mu+s\, p^\mu\,,
\end{equation}
where $s$ is some affine parameter and $p^\mu$ a constant null vector. A complete parametrisation of $p^\mu$ is further given in terms of three quantities $(\omega,w,\wb)$ by
\begin{equation}
\label{null pi}
p^\mu= \omega\, \hat q^\mu(w,\wb)\,,
\end{equation}
where $\hat q$ has been given in \eqref{q and n}. The components of $p^\mu$ in retarded coordinates along the null ray \eqref{null trajectory} are straightforwardly worked out,
\begin{align}
\begin{split}
p^r&=- n \cdot p=\omega\,,\\
p^u&=- \hat q \cdot p= \omega\, |z(s)-w|^2\,,\\
p^z&=r^{-1}\, \partial_{\zb}\hat q \cdot p=-\omega\, r(s)^{-1}\, (z(s)-w)\,.
\end{split}
\end{align}
Using \eqref{r and u} and \eqref{null trajectory} we find that the null trajectory is given in retarded coordinates by
\begin{align}
\begin{split}
r(s)&=-n \cdot X(s)=-n \cdot X_0+s\, \omega\,,\\
\label{u(s)}
u(s)&=-\hat q(z(s),\zb(s)) \cdot X(s)=-\hat q (z(s),\zb(s)) \cdot X_0+s\, \omega\, |z(s)-w|^2\,.
\end{split}
\end{align}
We note that $r(s)$ is an equally good affine parameter. The trajectory $z(s)$ can be obtained by studying the cartesian components $\mu=1,2$ of equation \eqref{coord transf retarded} and comparing it with \eqref{null trajectory}.\footnote{Note that $z(s)=w$ along the entire ray if the latter goes through the origin of Minkowski space, i.e., if $X_0=0$. Strictly speaking we need nonzero $X_0 \neq 0$ throughout the computations done in later sections, although we take the limit $X_0 \to 0$ at the very end with the simple effect of evaluating the soft fields at $u=0$.} For our purposes it enough to notice that in that limit $s \to \infty$, we have
\begin{equation}
\lim_{s \to \infty} z(s)=w\,,
\end{equation}
and therefore
\begin{equation}
\lim_{s \to \infty} u(s)=-\hat p \cdot X_0\,, \qquad \lim_{s \to \infty} \hat q^\mu(z(s),\zb(s))=\omega^{-1}\, p^\mu\,.
\end{equation}
Thus the null ray intersects $\scri^+$ at the retarded coordinates $(r,u,z,\zb)=(\infty,-\hat p \cdot X_0,w,\wb)$. 

\paragraph{Timelike infinity.}
Classical trajectories of massive particles asymptote to future timelike infinity $i^+$, which is better described using hyperbolic coordinates $(\tau,\rho,z,\zb)$, related to cartesian coordinates by \cite{Himwich:2020rro}
\begin{equation}
\label{coord transf hyperbolic}
X^\mu=\frac{1}{\sqrt{2}} \frac{\tau}{\rho} \left(n^\mu+\rho^2 \hat q^\mu(z,\zb) \right)\,,
\end{equation}
such that the flat metric takes the form
\begin{equation}
\label{metric hyperbolic}
ds^2=-d\tau^2+\tau^2 \left(\frac{d\rho^2}{\rho^2}+\rho^2 \dd z \dd \zb \right)\,.
\end{equation}
This is a foliation of the causal future of $X^\mu=0$ by three-dimensional hyperbolic slices of constant negative curvature (aka euclidean AdS$_3$) with coordinates $x^a=(\rho,z,\zb)$ and induced metric
\begin{equation}
\label{metric H}
ds^2_{\mathcal{H}}=h_{ab} \dd x^a \dd x^b=\frac{d\rho^2}{\rho^2}+\rho^2 \dd z \dd \zb\,.
\end{equation}
The asymptotic hyperbolic slice $\mathcal{H}^+$ located at $\tau \to \infty$ is a resolution of timelike infinity~$i^+$.
Note that we can easily extract the coordinates $(\tau,\rho)$ from the position vector $X^\mu$ through the following Lorentz contractions,
\begin{equation}
\tau^2=2\, (n \cdot X)\, (\hat q(z,\zb) \cdot X)\,, \qquad
\rho^2=\frac{n \cdot X}{\hat q(z,\zb) \cdot X}\,.
\end{equation}
By comparison of \eqref{coord transf hyperbolic} with \eqref{coord transf retarded}, we also easily infer the relation to retarded coordinates,
\begin{equation}
\label{retarded to hyperbolic}
u=\frac{1}{\sqrt{2}} \frac{\tau}{\rho}\,, \qquad r=\frac{1}{\sqrt{2}}\, \tau \rho\,.
\end{equation}
The limit $\rho \to \infty$ with $\tau$ fixed thus corresponds to the limit $r \to \infty\,, u \to 0$ towards the middle of $\scri^+$ where hyperbolic slices attach. The other limit of interest is $\tau \to \infty$ at fixed $\rho$ towards the future corner of $\scri^+$, which corresponds to $r\,,u \to \infty$. Finally the limit $r \to \infty$ with $u$ fixed but arbitrary corresponds to the limit $\rho\,, \tau \to \infty$ taken at the same rate.

A generic classical trajectory for a massive particle is still of the form \eqref{null trajectory}, with $p^\mu$ a constant timelike vector satisfying $p^2=-m^2$. We parametrise such momentum vectors by three numbers $(\chi,w,\wb)$,
\begin{equation}
\label{massive p cartesian}
p^\mu=\frac{m}{\sqrt{2} \chi} \left(n^\mu+\chi^2\, \hat q^\mu(w,\wb)\right)\equiv m\, \hat p^\mu\,,
\end{equation}
such that
\begin{equation}
n \cdot \hat p=-\frac{\chi}{\sqrt{2}}\,, \qquad \hat q(z,\zb) \cdot \hat p=-\frac{1}{\sqrt{2} \chi} \left(1+\chi^2 |z-w|^2 \right)\,.
\end{equation}
The components of the momentum \eqref{massive p cartesian} in hyperbolic coordinates and along the trajectory \eqref{null trajectory} are given by
\begin{align}
\begin{split}
\hat p^\rho&=\frac{1}{2 \tau(s) \chi}\left(\chi^2 - \rho(s)^2-\rho(s)^2 \chi^2 |z(s)-w|^2 \right)\,,\\
\hat p^\tau&=\frac{1}{2 \rho(s) \chi}\left(\chi^2 + \rho(s)^2+\rho(s)^2 \chi^2 |z(s)-w|^2 \right)\,,\\
\hat p^z&=-\frac{\chi}{\tau(s) \rho(s)} (z(s)-w)\,.
\end{split}
\end{align}
In the limit $s \to \infty$, the classical trajectory satisfies $\tau(s)\approx s$ together with
\begin{equation}
\lim_{s \to \infty} z(s)=w\,,\qquad \lim_{s \to \infty} \rho(s)=\chi\,.
\end{equation}
Thus the timelike trajectory intersects the asymptotic hyperbolic slice $\mathcal{H}^+$ at the coordinates $(\tau,\rho,z,\zb)=(\infty,\chi,w,\wb)$.

\section{Soft dressings of massless states}
\label{sec:massless soft dressings}
Starting from the Wilson line representations \eqref{leading Wilson}-\eqref{subleading Wilson} of the dressings, we aim at deriving their expressions in terms of asymptotic components of the photon field $A_\mu$. Our method will only reliably capture the soft part of the dressing since we will only retain the leading asymptotic contribution of the photon field. This is however all we need in order to capture the FK soft dressings as indicated by equation \eqref{soft W = tilde W}.

The field $A_\mu$ can be split into a radiative part and a piece sourced by charged on-shell particles. We will disregard the latter which is known to account for scattering Coulomb phases \cite{Jakob:1990zi}. The asymptotic expansion of the \textit{radiation} gauge field in retarded coordinates takes the form
\begin{equation}
\label{asymptotic expansion scri}
A_r=O(r^{-2} \ln r)\,, \qquad A_u=O(r^{-1} \ln r)\,, \qquad  A_z=A_z^{(0)}(u,z,\zb)+O(r^{-1})\,.
\end{equation}
Further information regarding this asymptotic expansion can be found in appendix~\ref{app:expansion at scri}. The asymptotic photon field admits the Fourier decomposition \cite{Strominger:2017zoo},
\begin{align}
A^{(0)}_z(u,z,\zb)=\frac{ie}{2\sqrt{2}\pi^2}\int_0^\infty d\omega\, ( a_+(q)\, e^{-i\omega u}-a^\dagger_-(q)\, e^{i\omega u} )\,, \qquad q=\omega \hat q(z,\zb)\,,
\end{align}
where $a^\dagger_+(q)$ is the usual creation operator for a positive helicity photon of momentum $q$. In the soft limit $\omega \to 0$ the two photon polarisations are however not independent, since they can be written in terms of the Goldstone mode $\Phi$ of large U(1) transformations,
\begin{align}
\label{a in terms of phi}
\begin{split}
a_+(\omega \hat q(z,\zb))&=\partial_z \Phi(\omega,z,\zb)\,, \qquad a_+^\dagger(\omega \hat q(z,\zb))=\partial_{\zb} \Phi^\dagger(\omega,z,\zb)\,, \qquad (\omega \to 0)\,,\\
a_-(\omega \hat q(z,\zb))&=\partial_{\zb} \Phi(\omega,z,\zb)\,, \qquad a_-^\dagger(\omega \hat q(z,\zb))=\partial_{z} \Phi^\dagger(\omega,z,\zb)\,.
\end{split}
\end{align}
We formally write the Goldstone mode as a soft field,
\begin{equation}
\label{Goldstone}
\Phi(u,z,\zb)=\frac{ie}{2\sqrt{2}\pi^2}\int_0^\infty d\omega\, f(\omega) \left( \Phi(\omega,z,\zb) e^{-i\omega u}-\Phi^\dagger(\omega,z,\zb) e^{i\omega u} \right)\,,
\end{equation}
such that
\begin{equation}
\label{A0 soft}
A^{(0)}_z(u,z,\zb)\big|_{\text{soft}}=\partial_z \Phi(u,z,\zb)\,.
\end{equation}
Furthermore one can construct holomorphic and antiholomorphic soft currents $S_z=\partial_z \Phi$ and $S_{\zb}=\partial_{\zb} \Phi$,  since away from operator insertions we have \cite{Nande:2017dba}
\begin{equation}
\partial_{\zb} \partial_z \Phi= 0\,.
\end{equation}
In other words the Goldstone mode can be decomposed into holomorphic and antiholomorphic components,
\begin{equation}
\label{small phis}
\Phi(u,z,\zb)=\phi(u,z)+\bar \phi(u,\zb)\,.
\end{equation}

The relations \eqref{a in terms of phi} then allow to write the soft components of the conformal fields \eqref{S and Y} in terms of the Goldstone mode and its velocity at the retarded time $u=0$,
\begin{equation}
\label{soft Goldstones}
\mathcal{S}\big|_{\text{soft}}=\Phi\big|_{u=0}\,, \qquad \partial_z \mathcal{Y}^z\big|_{\text{soft}}=\frac{1}{2} \partial_u \phi\big|_{u=0}\,, \qquad \partial_{\zb} \mathcal{Y}^{\zb}\big|_{\text{soft}}=\frac{1}{2} \partial_u \bar \phi\big|_{u=0}\,.
\end{equation}
This will come in handy when comparing our results to the conformally soft dressings \eqref{W0}-\eqref{W1}. 

\paragraph{Leading soft dressing.}
We have now introduced all the ingredients needed to evaluate the Wilson line dressing of a charged momentum state,
\begin{equation}
\exp \left[-i Q \int A\right] |p,J \rangle\,.
\end{equation}
Therefore the relevant quantity to compute is the line integral along the null ray \eqref{null trajectory},
\begin{align}
\label{int A}
\int A=\int_0^\infty ds\, p^\mu A_\mu(x(s))\,.
\end{align}
We will now make use of the asymptotic expansion \eqref{asymptotic expansion scri} of the radiation field $A_\mu$ near $\scri^+$. Since we are only interested in the leading soft contribution to the integral \eqref{int A}, it will be sufficient to consider the leading term of this asymptotic expansion. 

Plugging \eqref{asymptotic expansion scri} together with \eqref{A0 soft}, the leading soft contribution is therefore given by
\begin{equation}
\int_{0}^\infty ds\, p^\mu A_\mu(x(s))\Soft= \int_{0}^\infty ds \left(p^z \partial_z \Phi+ p^{\zb} \partial_{\zb} \Phi\right)= \int_{0}^\infty ds\, \frac{d}{ds} \Phi\,,
\end{equation}
where we have used 
\begin{equation}
\frac{d}{ds} \Phi(u(s),z(s),\zb(s))= p^z \partial_z \Phi+ p^{\zb} \partial_{\zb} \Phi+ p^u \partial_u \Phi \approx  p^z \partial_z \Phi+ p^{\zb} \partial_{\zb} \Phi\,.
\end{equation}
In the last equality we have discarded $\partial_u \Phi$ since it is subleading in the soft expansion ($\partial_u \sim \omega$). We thus have 
\begin{align}
\int A\, \Soft &=  \Phi(0,w,\wb) -\Phi_0 \,.
\end{align}
Just like in the general discussion \eqref{global phase}, $\Phi_0\equiv \Phi(0,z_0,\zb_0)$ is a phase common to all dressings irrespective of the insertion point $(w,\wb)$ on the celestial sphere and we can safely disregard it. Hence we find that the leading soft dressing of is simply given by
\begin{equation}
\tilde W_0=W_0\soft=e^{-iQ\, \Phi(0,w,\wb)}\,,
\end{equation}
in agreement with the soft part of the conformally soft dressing \eqref{W0} through the identification \eqref{soft Goldstones}. Note that the above expression had been considered previously to account for virtual IR divergences of the S-matrix in the approach of celestial holography \cite{Nande:2017dba}. Here we obtained it directly from the Wilson line dressing.

\paragraph{Subleading soft dressing.}
We similarly compute the subleading soft dressing starting from the generalised Wilson line,
\begin{equation}
\exp\left[-\frac{iQ}{8} \int ds\, F_{\mu\nu}(x(s)) J^{\mu\nu}\right] |p,J \rangle \,.
\label{def:sub}
\end{equation}
The asymptotic expansion \eqref{asymptotic expansion scri} together with \eqref{A0 soft} imply that the only nonvanishing soft components of the field strength are
\begin{equation}
F_{uA}\soft=\partial_u \partial_A \Phi\,,
\end{equation}
such that
\begin{equation}
\label{Fsoft J}
\begin{split}
F^{\mu\nu}\soft\, J_{\mu\nu}&=-2r^{-2} \left(F_{uz}\, J_{r\zb}+F_{u \zb}\, J_{rz} \right)=-2r^{-1} \left(F_{uz}\, \partial_{\zb} \hat q^\nu+F_{u \zb}\, \partial_z \hat q^\nu \right) \hat q^\mu J_{\mu\nu} \,.
\end{split}
\end{equation}
At this point it is convenient to express the momentum state parametrised by \eqref{null pi} in terms of conformal primary states \cite{Pasterski:2017kqt}, 
\begin{align}
|p, J \rangle= \int_{1-i \infty}^{1+i\infty} d\Delta\, \omega^{-\Delta}\, |\Delta,J,w,\wb \rangle\,, \qquad p^\mu=\omega\, \hat q^\mu(w,\wb)\,,
\end{align}
and exploit the action of the Lorentz generators \cite{Pasterski:2021dqe}
\begin{equation}
\hat q^\mu \partial_{\zb} \hat q^\nu J_{\mu\nu}\, |\Delta,J,w,\wb \rangle=-2i\left(2h(w-z)+(w-z)^2\partial_w \right)|\Delta,J,w,\wb \rangle\,,
\end{equation}
where the conformal weights are defined as
\begin{equation}
(h,\bar h)=\frac{1}{2}(\Delta+J,\Delta-J)\,.
\end{equation}
On a conformal primary state $|\Delta,J,w,\wb \rangle$, the action of \eqref{Fsoft J} therefore gives
\begin{equation}
\begin{split}
-i F^{\mu \nu}\soft\, J_{\mu \nu} &= 4 r^{-1} F_{uz}\left(2h(w-z)+(w-z)^2\partial_w \right)+\cc\\ 
&= 4 \omega^{-1} \left( 2 h p^z \partial_z \partial_u \Phi  + p^z \partial_z \partial_u \Phi (w-z) \partial_w + \cc   \right)\\
&=4 \omega^{-1} \left( 2 h  p^z \partial_z \partial_u \Phi+  p^z \partial_z[ \partial_u \Phi (w-z)]\partial_w +p^z \partial_u \Phi\, \partial_w + \cc \right)\,.
\end{split}
\end{equation}
Using the decomposition \eqref{small phis} together with the relations \eqref{soft Goldstones}, we can now rewrite this as a total derivative term,
\begin{equation}
\begin{split}
-iF^{\mu \nu}\soft\, J_{\mu \nu} &=8 \omega^{-1} p^z \partial_z \left[2 h   \partial_z \mathcal{Y}^z+  \partial_z\mathcal{Y}^z (w-z)\partial_w+ \mathcal{Y}^z\, \partial_w  \right]\soft+\cc\\
&=8 \omega^{-1} \frac{d}{ds} \left(2 h   \partial_z \mathcal{Y}^z+  \partial_z\mathcal{Y}^z (w-z)\partial_w+ \mathcal{Y}^z\, \partial_w  \right)\soft+\cc\,,
\end{split}
\end{equation}
such that 
\begin{equation}
W_1\soft\,  |p,J \rangle=\tilde W_1\, |p,J \rangle= \int_{1-i \infty}^{1+i\infty} d\Delta\, \omega^{-\Delta}\, \tilde W_1^{\text{conf}} |\Delta,J,w,\wb \rangle\,, 
\end{equation}
with 
\begin{equation}
\tilde W_1^{\text{conf}}=\exp\left[\frac{Q}{\omega} \left(2h\, \partial_w \mathcal{Y}^w+\mathcal{Y}^w \partial_w+2\bar h\, \partial_{\wb} \mathcal{Y}^{\wb}+\mathcal{Y}^{\wb} \partial_{\wb} \right)\soft\right]\,,
\end{equation}
in agreement with \eqref{W1} as given in \cite{Pasterski:2021dqe}.

We have thus recovered the soft contributions to the leading and subleading conformal dressings associated with charged massless states, starting from the generalised Wilson line operators.

\section{Soft dressings of massive states}
\label{sec:massive soft dressings}
Similarly to the case of massless fields worked out in the previous section, the determination of the soft dressings associated with charged massive states requires a control over the asymptotics of the radiation field $A_\mu$ near timelike infinity $i^+$. We will make use of the following expansion in the large $\tau$ limit,
\begin{align}
\label{large tau}
A_\tau=O(\tau^{-2})\,, \qquad A_a=\partial_a \Phi_{\mathcal{H}}+O(\tau^{-1})\,,
\end{align}
where $\Phi_{\mathcal{H}}$ is would-be pure gauge mode satisfying 
\begin{equation}
\label{3d box}
\square_{\mathcal{H}} \Phi_{\mathcal{H}}=0\,.
\end{equation}
Note that another scalar mode is in principle allowed by the asymptotic equations of motion in $A_\rho$ at order $O(\tau^0)$. However this mode is not produced by the standard free photon field operator and we therefore disregard it. Further details regarding this asymptotic expansion can be found in appendix~\ref{app:expansion at i+}. Solutions to \eqref{3d box} are fully determined in terms of the boundary value $\Phi^+(z,\zb)=\lim_{\rho \to \infty} \Phi_{\mathcal{H}}(\rho,z,\zb)$ \cite{Campiglia:2015qka,Nande:2017dba},
\begin{equation}
\label{Phi_H solution}
\Phi_{\mathcal{H}}(\rho,z,\zb)=\int d^2w\, K_2(\rho,z,\zb;w,\wb)\, \Phi^+(w,\wb)\,,
\end{equation}
where $K_2$ is a bulk-boundary propagator given in \eqref{K Delta}.

\paragraph{Leading dressing.} The determination of the leading soft dressing proceeds by evaluating the following Wilson line integral along the timelike trajectory \eqref{null trajectory} and \eqref{massive p cartesian}. Using the falloffs \eqref{large tau}, its soft contribution reduces to
\begin{equation}
\int dx^\mu A_\mu\Soft=\int ds\, p^a \partial_a  \Phi_{\mathcal{H}}=\int ds\, \frac{d}{ds} \Phi_{\mathcal{H}}=\Phi_{\mathcal{H}}(\chi,w,\wb)-\Phi_{\mathcal{H}}(\rho_0,z_0,\zb_0)\,.
\end{equation}
Again the second term is phase common to all dressings, and we can discard it. The leading soft conformal dressing of a massive particle is thus given by 
\begin{equation}
\tilde W_0=W_0\soft=e^{-iQ\, \Phi_{\mathcal{H}}(\chi,w,\wb)}\,,
\end{equation}
where $\Phi_{\mathcal{H}}$ is determined by \eqref{Phi_H solution}. In this way we recover the result of the literature \cite{Nande:2017dba,Arkani-Hamed:2020gyp}.

\paragraph{Subleading soft dressing.}
To determine the subleading soft dressing, we would need to evaluate the soft contribution to the line integral of $F_{\mu\nu} J^{\mu\nu}$. From the falloffs \eqref{large tau}, we see that the field strength vanishes near $i^+$,
\begin{equation}
F_{\tau a}=O(\tau^{-2})\,, \qquad F_{ab}=O(\tau^{-1})\,.
\end{equation}
Given that the leading soft dressing comes from the order $O(\tau^0)$ in the gauge potential, we can expect the subleading dressing to be associated with the subleading order $O(\tau^{-1})$. Asymptotically the time coordinate $\tau$ coincides with the Minkowskian time $t$ conjugated to the energy $\omega$. Therefore given a quantity $g(t)$ in the time-domain and its Fourier transform $g(\omega)$, we also have
\begin{equation}
\lim_{\tau \to \infty} g(\tau)=\lim_{\tau \to \infty} \int_0^\infty \dd\omega\, e^{-i\omega \tau} g(\omega)\,.
\end{equation}
The soft expansion of $g(\omega)$ in the limit $\omega \to 0$,
\begin{equation}
g(\omega)=\sum_{n=0} g_n\, \omega^n\,,
\end{equation}
then maps to an expansion at large time $\tau \to \infty$ through the Laplace transform
\begin{equation}
\int_0^\infty \dd \omega\, e^{-i\omega\tau}\, \omega^n=\frac{n!}{(i\tau)^{n+1}}\,.
\end{equation}
Thus higher orders in the soft expansion naturally map to higher orders the late-time asymptotic expansion towards $i^+$.

The explicit evaluation of the subleading soft dressing thus requires a detailed study of the Maxwell field near $i^+$ at subleading order $O(\tau^{-1})$ in the large-$\tau$ expansion. This can be done along the lines of appendix~\ref{app:expansion at i+}. In particular one needs to ensure consistency of the solution space near $i^+$ with that considered at $\scri^+$ in appendix~\ref{app:expansion at scri}. This however goes beyond the scope of the present article. Note that a similar systematic study of the solution space of general relativity near spatial infinity $i^0$ and its relation with the solution space near $\scri^+$ is discussed in \cite{Capone:2022gme}. As with the leading soft dressing, one can expect the subleading soft dressing of a massive state to resemble that of a massless state, modulo convolution by appropriate AdS$_3$ bulk-boundary propagators. We leave this to future endeavors.

\section*{Acknowledgments}
We thank Sangmin Choi, Ana Raclariu and Jakob Salzer for valuable discussions. The work of KN and CDW was supported by the UK Science and Technology Facilities Council (STFC) grants ST/P000258/1 and ST/T000759/1. The work of ARF was supported by the Royal society grant RF/ERE/210168 which is part of the Royal Society URF grant ``The Atoms of a de Sitter Universe''.

\appendix

\section{Asymptotic expansion near $\scri^+$}
\label{app:expansion at scri}
Here we discuss the asymptotics of the gauge field $A_\mu$ in retarded coordinates \eqref{flat Bondi gauge}, in close parallel to the analysis performed \cite{Himwich:2019dug}. Although in the main body of the text we restrict our attention to radiative solutions $(j_\mu=0)$, we keep the discussion general here.

Maxwell equations $\partial_\mu(\sqrt{-g} F^{\mu\nu})=e^2 \sqrt{-g}\, j^\nu$ take the form
\begin{align}
\begin{split}
-r^2 \partial_u F_{ru}+\partial_z F_{\zb u}+\partial_{\zb} F_{zu}&=e^2 r^2 j_u\,,\\
-\partial_r\left( r^2 F_{ur}\right)+\partial_z F_{\zb r}+\partial_{\zb} F_{zr}&=e^2 r^2 j_r\,,\\
-r^2 \partial_u F_{rz}- r^2\partial_r F_{uz}+\partial_z F_{\zb z}&=e^2 r^2 j_z\,,
\end{split}
\end{align}
while the Lorenz gauge condition $\partial_\mu(\sqrt{-g} A^\mu)=0$ reads
\begin{equation}
-r^2 \partial_u A_r-\partial_r\left( r^2 A_u\right)+\partial_z A_{\zb}+\partial_{\zb} A_z=0\,.
\label{eq:lorenz_g}
\end{equation}

We assume the following asymptotic expansion for the gauge field,\footnote{A nonzero $A^{(1)}_u$ can always be set to zero by a residual gauge transformation $A_\mu \mapsto A_\mu + \partial_\mu \varepsilon$ with 
\begin{equation}
\nonumber
\varepsilon(r,u,z,\zb)=r^{-1} \varepsilon^{(1)}(u,z,\zb)+O(r^{-2})\,, \qquad \partial_u \varepsilon^{(1)}=A_u^{(1)}\,.
\end{equation}}
\begin{align}
\begin{split}
A_r&=\sum_{n=2} r^{-n} A_r^{(n)}+\sum_{m=2} r^{-m} \ln r\, \tilde A_r^{(m)}\,,\\
A_u&=\sum_{n=2} r^{-n} A_u^{(n)}+\sum_{m=1} r^{-m} \ln r\, \tilde A_u^{(m)}\,,\\
A_z&=\sum_{n=0} r^{-n} A_z^{(n)}+\sum_{m=1} r^{-m} \ln r\, \tilde A_z^{(m)}\,,
\end{split}
\end{align}
and for the matter current,
\begin{equation}
j_u=r^{-2} j_u^{(2)}+O(r^{-3})\,, \qquad j_z=r^{-2}j_z^{(2)}+O(r^{-3})\,, \qquad j_r=O(r^{-3})\,.
\end{equation}
Under these assumptions the gauge condition \eqref{eq:lorenz_g} imposes
\begin{equation}
\label{gauge condition tilde}
\begin{split}
\p_u \tilde{A}_r^{(2)} &= -\tilde{A}_u^{(1)}\,, \\ 
\p_u A_r^{(2)}&= - \tilde{A}_u^{(1)}+\partial_{\zb}A_z^{(0)}+ \p_z A_{\zb}^{(0)}\,,
\end{split}
\end{equation}
while Maxwell equations additionally imply
\begin{equation}
\begin{split}
- 2\p_u \tilde{A}_u^{(1)} &= e^2 j_u^{(2)}\,, \\ 
\p_u \tilde{A}_z^{(2)}-\p_z \tilde{A}_u^{(1)} &= 0\,,\\ 
2 (\partial_z \partial_{\zb} A_z^{(0)} + \p_u A_z^{(1)})-2 \p_u \tilde{A}_z^{(1)} &=e^2 j_z^{(2)}\,.
\end{split}
\end{equation}
Although the coordinate system \eqref{flat Bondi gauge} is slightly different than the one used in \cite{Himwich:2019dug}, the equations take exactly the same form. 

\section{Asymptotic expansion near $i^+$}
\label{app:expansion at i+}
In this appendix we work out the asymptotics of the gauge field $A_\mu$ close to future timelike infinity $i^+$. This is done by adopting the hyperbolic coordinates $(\tau,x^a)=(\tau,\rho,z,\zb)$ and working in the limit $\tau \to \infty$. See \cite{Campiglia:2015qka,Campiglia:2015lxa,Campiglia:2019wxe} for relevant earlier work.

In the hyperbolic slicing \eqref{metric hyperbolic}, Maxwell equations take the form
\begin{equation}
\begin{split}
h^{ab} D_a F_{b\tau}&=e^2 \tau^2 j_\tau\,,\\
-\tau\, \partial_\tau(\tau F_{\tau b})+h^{ca}D_c F_{ab}&=e^2 \tau^2 j_b\,,
\end{split}
\end{equation}
where $D_a$ is the Levi-Civita connection associated with the three-dimensional metric $h_{ab}$ and where $F_{\tau a}\,, F_{ab}$ are viewed as (1,0)- and (2,0)-tensors on $\mathcal{H}$, respectively. On the other hand the Lorenz gauge condition reads
\begin{equation}
-\partial_\tau(\tau^3 A_\tau)+\tau\, h^{ab} D_a A_b=0\,.
\end{equation}

We start by assuming the standard falloffs for the gauge field
\begin{equation}
A_\tau=\tau^{-1}\, \bar A_\tau+O(\tau^{-2})\,, \qquad A_a=\bar A_a+O(\tau^{-1})\,,
\end{equation}
such that the asymptotics of the field strength are
\begin{equation}
F_{a\tau}=\tau^{-1}\, D_a \bar A_\tau+O(\tau^{-2})\,, \qquad F_{ab}= \bar F_{ab}+O(\tau^{-1})\,, \qquad \bar F_{ab}\equiv D_a \bar A_b-D_b \bar A_a\,,
\end{equation}
For the matter current on the other hand, we assume
\begin{align}
j_\tau=\tau^{-3}\, \bar j_\tau+O(\tau^{-2})\,, \qquad j_a=\tau^{-2}\, \bar j_a+O(\tau^{-3})\,,
\end{align}
At leading order Maxwell equations simply yield
\begin{equation}
\label{3d Maxwell}
D^2 \bar A_\tau=e^2\, \bar j_\tau\,, \qquad D^a \bar F_{ab}=e^2\, \bar j_b\,,
\end{equation}
while the Lorenz condition yields
\begin{equation}
\label{3d Lorenz}
D_a \bar A^a=2 \bar A_\tau\,.
\end{equation}
For radiative solutions we set $j_\mu=0$, and we will see that consistency with the falloffs at $\scri^+$ in that case requires $\bar A_\tau=0$. We are then effectively left with a lower dimensional Maxwell theory on $\mathcal{H}$ in Lorenz gauge.

The asymptotic equations \eqref{3d Maxwell}-\eqref{3d Lorenz} are equations on the euclidean hyperboloid $\mathcal{H}$. As is familiar from the AdS/CFT correspondence, we can fully characterize their solutions in terms of a `Fefferman--Graham' (FG) expansion at large $\rho$. For ease of notation we will use the following diagonal form of the euclidean metric,
\begin{equation}
ds^2_{\mathcal{H}}=\frac{\dd \rho^2}{\rho^2}+\rho^2\, \delta_{ij} \dd x^i \dd x^j\,,
\end{equation}
instead of the off-diagonal one \eqref{metric H}. Once the FG expansion of the solutions to \eqref{3d Maxwell}-\eqref{3d Lorenz} are worked out, we can look at their consistency with the falloffs assumed at $\scri^+$. Indeed the regime $\tau \gg \rho \gg 1$ corresponds to the regime $r \gg u \gg 1$, i.e., to the future corner of null infinity. By explicit coordinate transformation \eqref{retarded to hyperbolic}, we have
\begin{equation}
\label{i0 to scri}
\begin{split}
A_u&=\frac{1}{\sqrt{2}} \left(\rho A_\tau-\tau^{-1}\rho^2 A_\rho \right)=\frac{\tau^{-1}}{\sqrt{2}} \left(\rho \bar A_\tau-\rho^2 \bar A_\rho \right)+O(\tau^{-2})\,,\\
A_r&=\frac{1}{\sqrt{2}} \left(\rho^{-1} A_\tau+ \tau^{-1} A_\rho \right)=\frac{\tau^{-1}}{\sqrt{2}} \left(\rho^{-1} \bar A_\tau+ \bar A_\rho \right)+O(\tau^{-2})\,.  
\end{split}
\end{equation}
This will come in handy when working out the matching between quantities in the two coordinate systems.

\paragraph{FG expansion of $\bar A_\tau$.} 
The massless equation $D^2 \bar A_\tau=0$ explicitly becomes
\begin{equation}
\partial_\rho (\rho^3 \partial_\rho \bar A_\tau))+\rho^{-1} \partial^2 \bar A_\tau=0\,,
\end{equation}
with asymptotic solution
\begin{equation}
\label{FG Atau}
\bar A_\tau=A_\tau^{(0)}+\rho^{-2} \ln \rho\, \tilde A_\tau+\rho^{-2}\, A^{(2)}_\tau+...\,.
\end{equation}
The independent free data for this second order differential equation is $A^{(0)}_\tau$ and $A^{(2)}_\tau$. However regularity of the solution at $\rho=0$ discards half of the solution space, and therefore relates them in a nonlocal way via bulk-boundary propagators.

\paragraph{FG expansion for $\bar A_a$.} The equations \eqref{3d Maxwell}-\eqref{3d Lorenz}
can be written
\begin{equation}
\partial_\rho(\rho^3 \bar A_\rho)+\rho^{-1} \delta^{ij} \partial_i \bar A_j=2 \rho\, \bar A_\tau\,,
\end{equation}
and
\begin{equation}
\label{Arho eom}
\begin{split}
\partial_\rho ( \rho\, \partial_\rho(\rho^3 \bar A_\rho))+\partial^2 \bar A_\rho&=2\partial_\rho (\rho^2 \bar A_\tau)\,,\\
\partial_\rho(\rho\, \partial_\rho \bar A_i))+\rho^{-3} \partial^2 \bar A_i&=-2 \partial_i (\bar A_\rho-\rho^{-1} \bar A_\tau)\,.
\end{split}
\end{equation}
If $\bar A_\tau \neq 0$, the leading behavior of $\bar A_\rho$ is given by
\begin{equation}
\bar A_{\rho}=\rho^{-1}\, A^{(0)}_\tau+...\,,
\end{equation}
such that, together with \eqref{FG Atau} and  \eqref{i0 to scri}, we find
\begin{equation}
A_r=\sqrt{2}\, \tau^{-1}\rho^{-1}\, A_\tau^{(0)}+...=O(r^{-1}u^0)\,.
\end{equation}
Thus our assumption at $\scri^+$ that $A_r=O(r^{-2} \ln r)$ from appendix~\ref{app:expansion at scri} requires $A^{(0)}_\tau=0$ and thus $\bar A_\tau=0$ from regularity at $\rho=0$. In that case the FG expansion of $\bar A_a$ reduces to
\begin{equation}
\label{FG Aa}
\begin{split}
\bar A_{\rho}&=\rho^{-3} \ln \rho\, \tilde A_\rho + \rho^{-3}\, A^{(3)}_\rho+...\,,\\
\bar A_i&=\ln \rho\, \tilde A_i+A_i^{(0)}+...\,.
\end{split}
\end{equation}
Here $(\tilde A_\rho\,,A_\rho^{(3)})$ and $(\tilde A_i\,,A^{(0)}_i)$ are pairs of free data, that are again partially related by bulk-boundary propagators due to the requirement of regularity at $\rho=0$. Again our assumptions that $A_z=O(r^0)$ at $\scri^+$ requires to set $\tilde A_i=0$. One could naively think that regularity at $\rho=0$ implies $A_i^{(0)}=0$ as well, but this is not quite the case. Indeed would-be pure gauge solutions are still allowed,
\begin{equation}
\bar A_a=\partial_a \Phi_{\mathcal{H}}\,.
\end{equation}
These trivially satisfy the reduced Maxwell equations \eqref{3d Maxwell}, while the Lorenz gauge condition implies
\begin{equation}
\square_{\mathcal{H}}\Phi_{\mathcal{H}}=0\,,
\end{equation}
whose solutions admit the large-$\rho$ expansion
\begin{equation}
\label{FG Phi}
\Phi_{\mathcal{H}}=\Phi_{\mathcal{H}}^{(0)}+\rho^{-2} \ln \rho\, \tilde \Phi_{\mathcal{H}}+\rho^{-2}\, \Phi_{\mathcal{H}}^{(2)}+...\,, \qquad \tilde \Phi_{\mathcal{H}}=\frac{1}{2}\partial^2 \Phi^{(0)}_{\mathcal{H}}\,.
\end{equation}
Hence the corresponding gauge field reads
\begin{equation}
\begin{split}
\bar A_\rho&=-2 \rho^{-3} \ln \rho\, \tilde \Phi_{\mathcal{H}}+\rho^{-3}\, (\tilde \Phi_{\mathcal{H}}-2 \Phi_{\mathcal{H}}^{(2)} )+...\,,\\
\bar A_i&=\partial_i \Phi^{(0)}_{\mathcal{H}}+...\,.
\end{split}
\end{equation}
This forms the subset of the solutions \eqref{FG Aa} for which $\tilde A_i=0$. 

Equation \eqref{Arho eom} admits other solutions of $\bar A_\rho$ that are not pure gauge. With $\bar A_\tau=0$ and making the change of variable $\psi=\rho^2\bar A_\rho$, the latter takes the form 
\begin{equation}
\left(D^2+1\right)\psi=0\,, 
\end{equation}
which is the equation of a massive scalar right at the Breitenlohner--Freedman bound $m^2=-1$ \cite{Breitenlohner:1982bm,Breitenlohner:1982jf}.
Regularity at $\rho=0$ uniquely specifies it in terms of the boundary mode $\tilde A_\rho$,
\begin{equation}
\bar A_\rho(\rho,z,\zb)=\rho^{-2}\int d^2w\, K_1(\rho,z,\zb\,;w,\wb) \tilde A_\rho(w,\wb)\,,
\end{equation}
where the bulk-boundary propagator for general $\Delta$ is explicitly given by \cite{Freedman:1998tz}
\begin{equation}
\label{K Delta}
K_\Delta(\rho,z,\zb\,;w,\wb)=C_\Delta \left(\frac{\rho}{1+\rho^2|z-w|^2}\right)^\Delta\,.
\end{equation}

\paragraph{Summary.} By consistency with the falloff conditions at $\scri^+$, we conclude that the radiative solution space near $i^+$ admits the simple asymptotic expansion 
\begin{equation}
A_\tau=O(\tau^{-2})\,, \qquad A_a=\bar A_a+O(\tau^{-1})\,,
\end{equation}
with
\begin{equation}
\begin{split}
\bar A_\rho&=\rho^{-3} \ln \rho\, (\tilde A_\rho-\partial^2 \Phi^{(0)}_{\mathcal{H}})+O(\rho^{-3})\,,\\
\bar A_i&=\partial_i \Phi_{\mathcal{H}}^{(0)}+O(\rho^{-2}\ln \rho)\,.
\end{split}
\end{equation}
The functions $\Phi_{\mathcal{H}}^{(0)}(z,\zb)$ and $\tilde A_\rho(z,\zb)$ specify the two kinds of regular radiative solutions consistent with the aforementioned falloff conditions. In particular the solution specified by $\Phi_{\mathcal{H}}^{(0)}(z,\zb)$ is would-be pure gauge. Finally, the relations \eqref{i0 to scri} together with the coordinate change \eqref{retarded to hyperbolic} allow to map this radiative data to radiative data in retarded coordinates, namely
\begin{align}
\lim_{u \to \infty} A_z^{(0)}(u,z,\zb)=\partial_z \Phi_{\mathcal{H}}^{(0)}\,, \qquad \lim_{u \to \infty} \tilde A_u^{(1)}(u,z,\zb)=\frac{1}{4}(\tilde A_\rho-\partial^2 \Phi^{(0)}_{\mathcal{H}})\,.
\end{align}

\bibliography{bibl}

\providecommand{\href}[2]{#2}\begingroup\raggedright\begin{thebibliography}{10}

\bibitem{Mott}
N.F.~Mott, \emph{{On the influence of radiative forces on the scattering of
  electrons}}, {\emph{Proc. Camb. Phil. Soc.} {\bfseries 27} (1931) 255}.

\bibitem{Bloch:1937pw}
F.~Bloch and A.~Nordsieck, \emph{{Note on the Radiation Field of the
  electron}}, \href{https://doi.org/10.1103/PhysRev.52.54}{\emph{Phys. Rev.}
  {\bfseries 52} (1937) 54}.

\bibitem{Yennie:1961ad}
D.R.~Yennie, S.C.~Frautschi and H.~Suura, \emph{{The infrared divergence
  phenomena and high-energy processes}},
  \href{https://doi.org/10.1016/0003-4916(61)90151-8}{\emph{Annals Phys.}
  {\bfseries 13} (1961) 379}.

\bibitem{Kinoshita:1962ur}
T.~Kinoshita, \emph{{Mass singularities of Feynman amplitudes}},
  \href{https://doi.org/10.1063/1.1724268}{\emph{J. Math. Phys.} {\bfseries 3}
  (1962) 650}.

\bibitem{Lee:1964is}
T.D.~Lee and M.~Nauenberg, \emph{{Degenerate Systems and Mass Singularities}},
  \href{https://doi.org/10.1103/PhysRev.133.B1549}{\emph{Phys. Rev.} {\bfseries
  133} (1964) B1549}.

\bibitem{Weinberg:1965nx}
S.~Weinberg, \emph{{Infrared photons and gravitons}},
  \href{https://doi.org/10.1103/PhysRev.140.B516}{\emph{Phys. Rev.} {\bfseries
  140} (1965) B516}.

\bibitem{Chung:1965zza}
V.~Chung, \emph{{Infrared Divergence in Quantum Electrodynamics}},
  \href{https://doi.org/10.1103/PhysRev.140.B1110}{\emph{Phys. Rev.} {\bfseries
  140} (1965) B1110}.

\bibitem{Kibble:1968sfb}
T.W.B.~Kibble, \emph{{Coherent Soft-Photon States and Infrared Divergences. i.
  Classical Currents}}, \href{https://doi.org/10.1063/1.1664582}{\emph{J. Math.
  Phys.} {\bfseries 9} (1968) 315}.

\bibitem{Kibble:1968oug}
T.W.B.~Kibble, \emph{{Coherent soft-photon states and infrared divergences. ii.
  mass-shell singularities of green's functions}},
  \href{https://doi.org/10.1103/PhysRev.173.1527}{\emph{Phys. Rev.} {\bfseries
  173} (1968) 1527}.

\bibitem{Kibble:1968npb}
T.W.B.~Kibble, \emph{{Coherent soft-photon states and infrared divergences.
  iii. asymptotic states and reduction formulas}},
  \href{https://doi.org/10.1103/PhysRev.174.1882}{\emph{Phys. Rev.} {\bfseries
  174} (1968) 1882}.

\bibitem{Kibble:1968lka}
T.W.B.~Kibble, \emph{{Coherent soft-photon states and infrared divergences. iv.
  the scattering operator}},
  \href{https://doi.org/10.1103/PhysRev.175.1624}{\emph{Phys. Rev.} {\bfseries
  175} (1968) 1624}.

\bibitem{Kulish:1970ut}
P.P.~Kulish and L.D.~Faddeev, \emph{{Asymptotic conditions and infrared
  divergences in quantum electrodynamics}},
  \href{https://doi.org/10.1007/BF01066485}{\emph{Theor. Math. Phys.}
  {\bfseries 4} (1970) 745}.

\bibitem{Agarwal:2021ais}
N.~Agarwal, L.~Magnea, C.~Signorile-Signorile and A.~Tripathi, \emph{{The
  infrared structure of perturbative gauge theories}},
  \href{https://doi.org/10.1016/j.physrep.2022.10.001}{\emph{Phys. Rept.}
  {\bfseries 994} (2023) 1} [\href{https://arxiv.org/abs/2112.07099}{{\ttfamily
  2112.07099}}].

\bibitem{Dirac:1955uv}
P.A.M.~Dirac, \emph{{Gauge invariant formulation of quantum electrodynamics}},
  \href{https://doi.org/10.1139/p55-081}{\emph{Can. J. Phys.} {\bfseries 33}
  (1955) 650}.

\bibitem{Mandelstam:1962mi}
S.~Mandelstam, \emph{{Quantum electrodynamics without potentials}},
  \href{https://doi.org/10.1016/0003-4916(62)90232-4}{\emph{Annals Phys.}
  {\bfseries 19} (1962) 1}.

\bibitem{Zwanziger:1973mp}
D.~Zwanziger, \emph{{Field renormalization and reduction formula in quantum
  electrodynamics}},
  \href{https://doi.org/10.1103/PhysRevLett.30.934}{\emph{Phys. Rev. Lett.}
  {\bfseries 30} (1973) 934}.

\bibitem{Jakob:1990zi}
R.~Jakob and N.G.~Stefanis, \emph{{Path dependent phase factors and the
  infrared problem in QED}},
  \href{https://doi.org/10.1016/0003-4916(91)90277-F}{\emph{Annals Phys.}
  {\bfseries 210} (1991) 112}.

\bibitem{Strominger:2017zoo}
A.~Strominger, \emph{{Lectures on the Infrared Structure of Gravity and Gauge
  Theory}},  \href{https://arxiv.org/abs/1703.05448}{{\ttfamily 1703.05448}}.

\bibitem{Strominger:2013lka}
A.~Strominger, \emph{{Asymptotic Symmetries of Yang-Mills Theory}},
  \href{https://doi.org/10.1007/JHEP07(2014)151}{\emph{JHEP} {\bfseries 07}
  (2014) 151} [\href{https://arxiv.org/abs/1308.0589}{{\ttfamily 1308.0589}}].

\bibitem{He:2014cra}
T.~He, P.~Mitra, A.P.~Porfyriadis and A.~Strominger, \emph{{New Symmetries of
  Massless QED}}, \href{https://doi.org/10.1007/JHEP10(2014)112}{\emph{JHEP}
  {\bfseries 10} (2014) 112} [\href{https://arxiv.org/abs/1407.3789}{{\ttfamily
  1407.3789}}].

\bibitem{Kapec:2015ena}
D.~Kapec, M.~Pate and A.~Strominger, \emph{{New Symmetries of QED}},
  \href{https://doi.org/10.4310/ATMP.2017.v21.n7.a7}{\emph{Adv. Theor. Math.
  Phys.} {\bfseries 21} (2017) 1769}
  [\href{https://arxiv.org/abs/1506.02906}{{\ttfamily 1506.02906}}].

\bibitem{Campiglia:2015qka}
M.~Campiglia and A.~Laddha, \emph{{Asymptotic symmetries of QED and
  Weinberg\textquoteright{}s soft photon theorem}},
  \href{https://doi.org/10.1007/JHEP07(2015)115}{\emph{JHEP} {\bfseries 07}
  (2015) 115} [\href{https://arxiv.org/abs/1505.05346}{{\ttfamily
  1505.05346}}].

\bibitem{Lysov:2014csa}
V.~Lysov, S.~Pasterski and A.~Strominger, \emph{{Low\textquoteright{}s
  Subleading Soft Theorem as a Symmetry of QED}},
  \href{https://doi.org/10.1103/PhysRevLett.113.111601}{\emph{Phys. Rev. Lett.}
  {\bfseries 113} (2014) 111601}
  [\href{https://arxiv.org/abs/1407.3814}{{\ttfamily 1407.3814}}].

\bibitem{Gabai:2016kuf}
B.~Gabai and A.~Sever, \emph{{Large gauge symmetries and asymptotic states in
  QED}}, \href{https://doi.org/10.1007/JHEP12(2016)095}{\emph{JHEP} {\bfseries
  12} (2016) 095} [\href{https://arxiv.org/abs/1607.08599}{{\ttfamily
  1607.08599}}].

\bibitem{Kapec:2017tkm}
D.~Kapec, M.~Perry, A.-M.~Raclariu and A.~Strominger, \emph{{Infrared
  Divergences in QED, Revisited}},
  \href{https://doi.org/10.1103/PhysRevD.96.085002}{\emph{Phys. Rev. D}
  {\bfseries 96} (2017) 085002}
  [\href{https://arxiv.org/abs/1705.04311}{{\ttfamily 1705.04311}}].

\bibitem{Choi:2017ylo}
S.~Choi and R.~Akhoury, \emph{{BMS Supertranslation Symmetry Implies
  Faddeev-Kulish Amplitudes}},
  \href{https://doi.org/10.1007/JHEP02(2018)171}{\emph{JHEP} {\bfseries 02}
  (2018) 171} [\href{https://arxiv.org/abs/1712.04551}{{\ttfamily
  1712.04551}}].

\bibitem{Low:1954kd}
F.E.~Low, \emph{{Scattering of light of very low frequency by systems of spin
  1/2}}, \href{https://doi.org/10.1103/PhysRev.96.1428}{\emph{Phys. Rev.}
  {\bfseries 96} (1954) 1428}.

\bibitem{Gell-Mann:1954wra}
M.~Gell-Mann and M.L.~Goldberger, \emph{{Scattering of low-energy photons by
  particles of spin 1/2}},
  \href{https://doi.org/10.1103/PhysRev.96.1433}{\emph{Phys. Rev.} {\bfseries
  96} (1954) 1433}.

\bibitem{Low:1958sn}
F.E.~Low, \emph{{Bremsstrahlung of very low-energy quanta in elementary
  particle collisions}},
  \href{https://doi.org/10.1103/PhysRev.110.974}{\emph{Phys. Rev.} {\bfseries
  110} (1958) 974}.

\bibitem{Burnett:1967km}
T.H.~Burnett and N.M.~Kroll, \emph{{Extension of the low soft photon theorem}},
  \href{https://doi.org/10.1103/PhysRevLett.20.86}{\emph{Phys. Rev. Lett.}
  {\bfseries 20} (1968) 86}.

\bibitem{Choi:2019rlz}
S.~Choi and R.~Akhoury, \emph{{Subleading soft dressings of asymptotic states
  in QED and perturbative quantum gravity}},
  \href{https://doi.org/10.1007/JHEP09(2019)031}{\emph{JHEP} {\bfseries 09}
  (2019) 031} [\href{https://arxiv.org/abs/1907.05438}{{\ttfamily
  1907.05438}}].

\bibitem{Laenen:2008gt}
E.~Laenen, G.~Stavenga and C.D.~White, \emph{{Path integral approach to eikonal
  and next-to-eikonal exponentiation}},
  \href{https://doi.org/10.1088/1126-6708/2009/03/054}{\emph{JHEP} {\bfseries
  03} (2009) 054} [\href{https://arxiv.org/abs/0811.2067}{{\ttfamily
  0811.2067}}].

\bibitem{Laenen:2010uz}
E.~Laenen, L.~Magnea, G.~Stavenga and C.D.~White, \emph{{Next-to-Eikonal
  Corrections to Soft Gluon Radiation: A Diagrammatic Approach}},
  \href{https://doi.org/10.1007/JHEP01(2011)141}{\emph{JHEP} {\bfseries 01}
  (2011) 141} [\href{https://arxiv.org/abs/1010.1860}{{\ttfamily 1010.1860}}].

\bibitem{White:2014qia}
C.D.~White, \emph{{Diagrammatic insights into next-to-soft corrections}},
  \href{https://doi.org/10.1016/j.physletb.2014.08.041}{\emph{Phys. Lett. B}
  {\bfseries 737} (2014) 216}
  [\href{https://arxiv.org/abs/1406.7184}{{\ttfamily 1406.7184}}].

\bibitem{Bahjat-Abbas:2019fqa}
N.~Bahjat-Abbas, D.~Bonocore, J.~Sinninghe~Damst\'e, E.~Laenen, L.~Magnea,
  L.~Vernazza et~al., \emph{{Diagrammatic resummation of leading-logarithmic
  threshold effects at next-to-leading power}},
  \href{https://doi.org/10.1007/JHEP11(2019)002}{\emph{JHEP} {\bfseries 11}
  (2019) 002} [\href{https://arxiv.org/abs/1905.13710}{{\ttfamily
  1905.13710}}].

\bibitem{White:2011yy}
C.D.~White, \emph{{Factorization Properties of Soft Graviton Amplitudes}},
  \href{https://doi.org/10.1007/JHEP05(2011)060}{\emph{JHEP} {\bfseries 05}
  (2011) 060} [\href{https://arxiv.org/abs/1103.2981}{{\ttfamily 1103.2981}}].

\bibitem{Bonocore:2015esa}
D.~Bonocore, E.~Laenen, L.~Magnea, S.~Melville, L.~Vernazza and C.D.~White,
  \emph{{A factorization approach to next-to-leading-power threshold
  logarithms}}, \href{https://doi.org/10.1007/JHEP06(2015)008}{\emph{JHEP}
  {\bfseries 06} (2015) 008}
  [\href{https://arxiv.org/abs/1503.05156}{{\ttfamily 1503.05156}}].

\bibitem{Bonocore:2016awd}
D.~Bonocore, E.~Laenen, L.~Magnea, L.~Vernazza and C.D.~White,
  \emph{{Non-abelian factorisation for next-to-leading-power threshold
  logarithms}}, \href{https://doi.org/10.1007/JHEP12(2016)121}{\emph{JHEP}
  {\bfseries 12} (2016) 121}
  [\href{https://arxiv.org/abs/1610.06842}{{\ttfamily 1610.06842}}].

\bibitem{Pasterski:2021raf}
S.~Pasterski, M.~Pate and A.-M.~Raclariu, \emph{{Celestial Holography}},  in
  \emph{{2022 Snowmass Summer Study}}, 11, 2021
  [\href{https://arxiv.org/abs/2111.11392}{{\ttfamily 2111.11392}}].

\bibitem{McLoughlin:2022ljp}
T.~McLoughlin, A.~Puhm and A.-M.~Raclariu, \emph{{The SAGEX Review on
  Scattering Amplitudes, Chapter 11: Soft Theorems and Celestial Amplitudes}},
  \href{https://arxiv.org/abs/2203.13022}{{\ttfamily 2203.13022}}.

\bibitem{deBoer:2003vf}
J.~de~Boer and S.N.~Solodukhin, \emph{{A Holographic reduction of Minkowski
  space-time}},
  \href{https://doi.org/10.1016/S0550-3213(03)00494-2}{\emph{Nucl. Phys.}
  {\bfseries B665} (2003) 545}
  [\href{https://arxiv.org/abs/hep-th/0303006}{{\ttfamily hep-th/0303006}}].

\bibitem{Cheung:2016iub}
C.~Cheung, A.~de~la Fuente and R.~Sundrum, \emph{{4D scattering amplitudes and
  asymptotic symmetries from 2D CFT}},
  \href{https://doi.org/10.1007/JHEP01(2017)112}{\emph{JHEP} {\bfseries 01}
  (2017) 112} [\href{https://arxiv.org/abs/1609.00732}{{\ttfamily
  1609.00732}}].

\bibitem{Pasterski:2016qvg}
S.~Pasterski, S.-H.~Shao and A.~Strominger, \emph{{Flat Space Amplitudes and
  Conformal Symmetry of the Celestial Sphere}},
  \href{https://doi.org/10.1103/PhysRevD.96.065026}{\emph{Phys. Rev. D}
  {\bfseries 96} (2017) 065026}
  [\href{https://arxiv.org/abs/1701.00049}{{\ttfamily 1701.00049}}].

\bibitem{Pasterski:2017kqt}
S.~Pasterski and S.-H.~Shao, \emph{{Conformal basis for flat space
  amplitudes}}, \href{https://doi.org/10.1103/PhysRevD.96.065022}{\emph{Phys.
  Rev. D} {\bfseries 96} (2017) 065022}
  [\href{https://arxiv.org/abs/1705.01027}{{\ttfamily 1705.01027}}].

\bibitem{Pasterski:2017ylz}
S.~Pasterski, S.-H.~Shao and A.~Strominger, \emph{{Gluon Amplitudes as 2d
  Conformal Correlators}},
  \href{https://doi.org/10.1103/PhysRevD.96.085006}{\emph{Phys. Rev. D}
  {\bfseries 96} (2017) 085006}
  [\href{https://arxiv.org/abs/1706.03917}{{\ttfamily 1706.03917}}].

\bibitem{Fotopoulos:2019vac}
A.~Fotopoulos, S.~Stieberger, T.R.~Taylor and B.~Zhu, \emph{{Extended BMS
  Algebra of Celestial CFT}},
  \href{https://doi.org/10.1007/JHEP03(2020)130}{\emph{JHEP} {\bfseries 03}
  (2020) 130} [\href{https://arxiv.org/abs/1912.10973}{{\ttfamily
  1912.10973}}].

\bibitem{Law:2020tsg}
Y.T.A.~Law and M.~Zlotnikov, \emph{{Massive Spinning Bosons on the Celestial
  Sphere}}, \href{https://doi.org/10.1007/JHEP06(2020)079}{\emph{JHEP}
  {\bfseries 06} (2020) 079}
  [\href{https://arxiv.org/abs/2004.04309}{{\ttfamily 2004.04309}}].

\bibitem{Iacobacci:2020por}
L.~Iacobacci and W.~M\"uck, \emph{{Conformal Primary Basis for Dirac Spinors}},
  \href{https://doi.org/10.1103/PhysRevD.102.106025}{\emph{Phys. Rev. D}
  {\bfseries 102} (2020) 106025}
  [\href{https://arxiv.org/abs/2009.02938}{{\ttfamily 2009.02938}}].

\bibitem{Narayanan:2020amh}
S.A.~Narayanan, \emph{{Massive Celestial Fermions}},
  \href{https://doi.org/10.1007/JHEP12(2020)074}{\emph{JHEP} {\bfseries 12}
  (2020) 074} [\href{https://arxiv.org/abs/2009.03883}{{\ttfamily
  2009.03883}}].

\bibitem{Pasterski:2020pdk}
S.~Pasterski and A.~Puhm, \emph{{Shifting spin on the celestial sphere}},
  \href{https://doi.org/10.1103/PhysRevD.104.086020}{\emph{Phys. Rev. D}
  {\bfseries 104} (2021) 086020}
  [\href{https://arxiv.org/abs/2012.15694}{{\ttfamily 2012.15694}}].

\bibitem{He:2015zea}
T.~He, P.~Mitra and A.~Strominger, \emph{{2D Kac-Moody Symmetry of 4D
  Yang-Mills Theory}},
  \href{https://doi.org/10.1007/JHEP10(2016)137}{\emph{JHEP} {\bfseries 10}
  (2016) 137} [\href{https://arxiv.org/abs/1503.02663}{{\ttfamily
  1503.02663}}].

\bibitem{Himwich:2019dug}
E.~Himwich and A.~Strominger, \emph{{Celestial current algebra from
  Low\textquoteright{}s subleading soft theorem}},
  \href{https://doi.org/10.1103/PhysRevD.100.065001}{\emph{Phys. Rev. D}
  {\bfseries 100} (2019) 065001}
  [\href{https://arxiv.org/abs/1901.01622}{{\ttfamily 1901.01622}}].

\bibitem{Nande:2017dba}
A.~Nande, M.~Pate and A.~Strominger, \emph{{Soft Factorization in QED from 2D
  Kac-Moody Symmetry}},
  \href{https://doi.org/10.1007/JHEP02(2018)079}{\emph{JHEP} {\bfseries 02}
  (2018) 079} [\href{https://arxiv.org/abs/1705.00608}{{\ttfamily
  1705.00608}}].

\bibitem{Arkani-Hamed:2020gyp}
N.~Arkani-Hamed, M.~Pate, A.-M.~Raclariu and A.~Strominger, \emph{{Celestial
  amplitudes from UV to IR}},
  \href{https://doi.org/10.1007/JHEP08(2021)062}{\emph{JHEP} {\bfseries 08}
  (2021) 062} [\href{https://arxiv.org/abs/2012.04208}{{\ttfamily
  2012.04208}}].

\bibitem{Pasterski:2021dqe}
S.~Pasterski, A.~Puhm and E.~Trevisani, \emph{{Revisiting the conformally soft
  sector with celestial diamonds}},
  \href{https://doi.org/10.1007/JHEP11(2021)143}{\emph{JHEP} {\bfseries 11}
  (2021) 143} [\href{https://arxiv.org/abs/2105.09792}{{\ttfamily
  2105.09792}}].

\bibitem{Choi:2018oel}
S.~Choi and R.~Akhoury, \emph{{Soft Photon Hair on Schwarzschild Horizon from a
  Wilson Line Perspective}},
  \href{https://doi.org/10.1007/JHEP12(2018)074}{\emph{JHEP} {\bfseries 12}
  (2018) 074} [\href{https://arxiv.org/abs/1809.03467}{{\ttfamily
  1809.03467}}].

\bibitem{Duary:2022afn}
S.~Duary, \emph{{AdS correction to the Faddeev-Kulish state: migrating from the
  flat peninsula}}, \href{https://doi.org/10.1007/JHEP05(2023)079}{\emph{JHEP}
  {\bfseries 05} (2023) 079}
  [\href{https://arxiv.org/abs/2212.09509}{{\ttfamily 2212.09509}}].

\bibitem{Himwich:2020rro}
E.~Himwich, S.A.~Narayanan, M.~Pate, N.~Paul and A.~Strominger, \emph{{The Soft
  $\mathcal{S}$-Matrix in Gravity}},
  \href{https://doi.org/10.1007/JHEP09(2020)129}{\emph{JHEP} {\bfseries 09}
  (2020) 129} [\href{https://arxiv.org/abs/2005.13433}{{\ttfamily
  2005.13433}}].

\bibitem{Capone:2022gme}
F.~Capone, K.~Nguyen and E.~Parisini, \emph{{Charge and antipodal matching
  across spatial infinity}},
  \href{https://doi.org/10.21468/SciPostPhys.14.2.014}{\emph{SciPost Phys.}
  {\bfseries 14} (2023) 014}
  [\href{https://arxiv.org/abs/2204.06571}{{\ttfamily 2204.06571}}].

\bibitem{Campiglia:2015lxa}
M.~Campiglia, \emph{{Null to time-like infinity Green\textquoteright{}s
  functions for asymptotic symmetries in Minkowski spacetime}},
  \href{https://doi.org/10.1007/JHEP11(2015)160}{\emph{JHEP} {\bfseries 11}
  (2015) 160} [\href{https://arxiv.org/abs/1509.01408}{{\ttfamily
  1509.01408}}].

\bibitem{Campiglia:2019wxe}
M.~Campiglia and A.~Laddha, \emph{{Loop Corrected Soft Photon Theorem as a Ward
  Identity}}, \href{https://doi.org/10.1007/JHEP10(2019)287}{\emph{JHEP}
  {\bfseries 10} (2019) 287}
  [\href{https://arxiv.org/abs/1903.09133}{{\ttfamily 1903.09133}}].

\bibitem{Breitenlohner:1982bm}
P.~Breitenlohner and D.Z.~Freedman, \emph{{Positive Energy in anti-De Sitter
  Backgrounds and Gauged Extended Supergravity}},
  \href{https://doi.org/10.1016/0370-2693(82)90643-8}{\emph{Phys. Lett. B}
  {\bfseries 115} (1982) 197}.

\bibitem{Breitenlohner:1982jf}
P.~Breitenlohner and D.Z.~Freedman, \emph{{Stability in Gauged Extended
  Supergravity}},
  \href{https://doi.org/10.1016/0003-4916(82)90116-6}{\emph{Annals Phys.}
  {\bfseries 144} (1982) 249}.

\bibitem{Freedman:1998tz}
D.Z.~Freedman, S.D.~Mathur, A.~Matusis and L.~Rastelli, \emph{{Correlation
  functions in the CFT(d) / AdS(d+1) correspondence}},
  \href{https://doi.org/10.1016/S0550-3213(99)00053-X}{\emph{Nucl. Phys. B}
  {\bfseries 546} (1999) 96}
  [\href{https://arxiv.org/abs/hep-th/9804058}{{\ttfamily hep-th/9804058}}].

\end{thebibliography}\endgroup
\bibliographystyle{JHEP}  
\end{document}